\documentclass[conference,a4paper]{IEEEtran}

\newif\ifthreevalue

\usepackage[utf8]{inputenc} 
\usepackage[T1]{fontenc}
\usepackage{url}              
\usepackage[noadjust]{cite}             

\usepackage[cmex10]{amsmath}  
\usepackage{mleftright}       
\mleftright                   

\usepackage{graphicx}         
\usepackage{booktabs}         

\IEEEoverridecommandlockouts
\usepackage{amsmath,amssymb,amsfonts,amsthm}
\usepackage{enumitem}
\usepackage{mathtools}
\usepackage{algorithmic}
\usepackage{textcomp}
\usepackage[table,xcdraw]{xcolor}
\usepackage{tikz}

\usepackage{subcaption}
\usepackage{multicol,multirow}

\newtheorem{theorem}{Theorem}

\newtheorem{example}{Example}
\newtheorem{lemma}{Lemma}

\newtheorem{construction}{Construction}
\newtheorem{corollary}{Corollary}
\newtheorem{question}{Question}

\DeclareMathOperator*{\capacity}{cap}
\DeclareMathOperator*{\ocapacity}{\overline{cap}}

\newcommand{\cC}{\mathcal{C}}

\newcommand{\cH}{\mathcal{H}}

\newcommand{\cP}{\mathcal{P}}

\DeclarePairedDelimiter\abs{\lvert}{\rvert}
\DeclarePairedDelimiter\norm{\lVert}{\rVert}
\DeclarePairedDelimiter\ceil{\lceil}{\rceil}
\DeclarePairedDelimiter\floor{\lfloor}{\rfloor}
\DeclarePairedDelimiter\parenv{\lparen}{\rparen}

\DeclarePairedDelimiter\set{\{}{\}}

\newcommand{\F}{\mathbb{F}}
\newcommand{\eqdef}{\triangleq}
\renewcommand{\geq}{\geqslant}

\renewcommand{\leq}{\leqslant}
\renewcommand{\le}{\leqslant}

\DeclareFontFamily{U}{mathb}{}
\DeclareFontShape{U}{mathb}{m}{n}{
  <-5.5> mathb5
  <5.5-6.5> mathb6
  <6.5-7.5> mathb7
  <7.5-8.5> mathb8
  <8.5-9.5> mathb9
  <9.5-11.5> mathb10
  <11.5-> mathb12
}{}
\DeclareRobustCommand{\sqcdot}{%
  \mathbin{\text{\usefont{U}{mathb}{m}{n}\symbol{"0D}}}%
}
\newcommand{\bx}{{\sqcdot}}

\DeclareSymbolFont{bbold}{U}{bbold}{m}{n}
\DeclareSymbolFontAlphabet{\mathbbold}{bbold}

\begin{document}
\title{Bounds on Box Codes}
\author{
  \IEEEauthorblockN{ Michael Langberg\IEEEauthorrefmark{1}, Moshe Schwartz\IEEEauthorrefmark{2}\IEEEauthorrefmark{3}, and Itzhak Tamo\IEEEauthorrefmark{4}}
  \IEEEauthorblockA{\IEEEauthorrefmark{1} Department of Electrical Engineering, University at Buffalo, Buffalo, NY 14260, USA}
  \IEEEauthorblockA{\IEEEauthorrefmark{2} Department of Electrical and Computer Engineering, McMaster University, Hamilton, ON L8S 4K1, Canada}
  \IEEEauthorblockA{\IEEEauthorrefmark{3} School of Electrical and Computer Engineering, Ben-Gurion University of the Negev, Beer Sheva 8410501, Israel}
  \IEEEauthorblockA{\IEEEauthorrefmark{4} Department of Electrical Engineering--Systems, Tel Aviv University, Tel Aviv 6997801, Israel}
 \IEEEauthorblockA{Emails: \texttt{mikel@buffalo.edu}, \texttt{schwartz.moshe@mcmaster.ca}, \texttt{tamo@tauex.tau.ac.il}}
 \thanks{This work was supported in part by NSF grant CCF-2245204 and  by the European Research Council (ERC grant number 852953).}
}

\maketitle

\begin{abstract}
    Let $n_q(M,d)$ be the minimum length of a $q$-ary code of size $M$ and minimum distance $d$.
Bounding $n_q(M,d)$ is a fundamental problem that lies at the heart of coding theory.
This work considers a generalization $n^\bx_q(M,d)$ of $n_q(M,d)$ corresponding to codes in which codewords have \emph{protected} and \emph{unprotected} entries; where (analogs of) distance and of length are  measured with respect to protected entries only.
Such codes, here referred to as \emph{box codes}, have seen prior studies in the context of bipartite graph covering. 
Upper and lower bounds on $n^\bx_q(M,d)$ are presented.
\end{abstract}

\section{Introduction}

Let $n_q(M,d)$ be the minimum length of a $q$-ary code of size $M$ and minimum distance at least $d$.
Bounding $n_q(M,d)$ is a fundamental problem that lies at the heart of coding theory~\cite{MacSlo78}.
In this work, we consider a generalization of the functional $n_q(M,d)$ corresponding to codes in which  codewords have \emph{protected} and \emph{unprotected} entries.
We refer to such codes as \emph{box codes}. 
Specifically, for an alphabet $\Sigma_q$ of size $q$, a box code $\cC^\bx$ over $\Sigma_q$ consists of $M$ infinite-length codewords with entries in $\Sigma_q\cup\set{\bx}$.
Here, the symbol $\bx$ represents an \emph{unprotected} codeword entry -- in the sense that it may be received as any symbol from $\Sigma_q$, even without errors in the channel; symbols in $\Sigma_q$ represent \emph{protected} entries. Thus, the set of all words over $\Sigma_q$ that may be received from a given codeword, without any further errors introduced by the channel, forms a box in $\Sigma_q^\infty$.

One requires that codewords of a box code have only a finite number of protected entries from $\Sigma_q$. The \emph{length} of a codeword equals the number of protected entries it has. The length of a box code is the average length of its codewords. Additionally, the distance between two codewords equals the number of positions containing differing protected symbols. We can then ask what is the minimum length of a box code given the number of codewords it contains, and the minimum distance between its codewords.

We follow with a formal introduction to the main concepts that we study. Let $\Sigma_q$ be a finite alphabet of size $q$. A traditional code, $\cC$ with parameters $(n,M,d)_q$, is a subset $\cC\subseteq \Sigma_q^n$, $\abs{\cC}=M$, such that
\[
d = \min_{\substack{c,c'\in\cC \\ c\neq c'}} d(c,c'),
\]
where $d(c,c')$ is the Hamming distance between $c$ and $c'$, that equals the number of positions in which they disagree. If $\Sigma_q=\F_q$, the finite field of size $q$, and the code is a linear subspace of $\F_q^n$, we say the code is \emph{linear} and has parameter $[n,k,d]_q$, where $k=\log_q M$ is the dimension of $\cC$ as a vector space.

To define box codes, our alphabet needs an extra \emph{unprotected} symbol, $\bx$, and we define 
\[\Sigma \eqdef \Sigma_q \cup\set{\bx}.\] 
An infinite sequence $w=(w_1,w_2,\dots)\in\Sigma^\infty$ is called a \emph{word}. The \emph{length} of $w$, denoted $\norm{w}_\bx$, is the number of protected entries in $w$, namely,
\[
\norm*{w}_\bx \eqdef \abs*{\set*{i\geq 1 ~:~ w_i\in\Sigma_q}}.
\]
We note that $\norm{w}_\bx$ may be infinite in general.

Given two words, $w,w'\in\Sigma^\infty$, the \emph{distance} between them is defined as
\[
d^\bx(w,w') \eqdef \abs*{\set*{ i\geq 1 ~:~ w_i,w'_i\in\Sigma_q, w_i\neq w'_i}}.
\]

A box code $\cC^\bx$ is simply a finite subset $\cC^\bx\subseteq \Sigma^\infty$, that contains only words of finite length. The words of $\cC^\bx$ are called \emph{codewords}. We say $\cC^\bx$ has \emph{size} $M$ if $\abs{\cC^\bx}=M$. The \emph{length} of $\cC^\bx$, denoted $n$, is defined as the average length of its codewords,
\begin{equation}
\label{eq:length}
n \eqdef \frac{1}{M}\sum_{c\in\cC^\bx} \norm*{c}_\bx.
\end{equation}

We also define the \emph{minimum distance} of $\cC^\bx$, denoted $d$, as the smallest $d^\bx$-distance between distinct codewords, i.e.,
\[
d \eqdef \min_{\substack{c,c'\in\cC^\bx\\ c\neq c'}} d^\bx(c,c').
\]
We summarize all this information by saying that $\cC^\bx$ is an $(n,M,d)_q$ box code. Finally, we use $n^\bx_q(M,d)$ to denote the minimum length for which there exists a $q$-ary box code of size $M$ and minimum distance at least $d$. Notice that $n^\bx_q(M,d)$ need not be an integer.

Some remarks are in place.
First, notice that box codes generalize the standard notion of codes. Indeed, assume $\cC$ is an $(n,M,d)_q$ code over $\Sigma_q$. Thus, all its codewords are vectors of length $n$ over the protected symbols $\Sigma_q$. By concatenating $\bx^\infty$ to each codeword in $\cC$ we obtain $\cC^\bx$ that is an $(n,M,d)_q$ box code. This implies that $n_q^\bx(M,d) \leq n_q(M,d)$.

Moreover, since we are mainly interested in $n^\bx_q(M,d)$, we may reorder the coordinates of a box code $\cC^\bx$ to get an equivalent code. We therefore assume, without loss of generality, that there exists an integer $\eta$ such that all protected entries of $\cC^\bx$ are in the first $\eta$ coordinates, and $\eta$ is the minimal with this property. We call $\eta$ the \emph{protected length} of the box code. Note that in an $(n,M,d)_q$ box code $\cC^\bx$, the codewords have a protected length of at most $Mn$. This follows from the fact that~\eqref{eq:length} implies that the total number of protected symbols in the code is $Mn$. Thus, $\eta\leq Mn$. If the first $\eta$ coordinates of $\cC^\bx$ are all protected, the code is simply a traditional code concatenated with $\bx^\infty$ as discussed above, in which case we say $\cC^\bx$ is a \emph{degenerate} box code.

From an operational perspective, box codes lend themselves naturally to wireless communication settings, e.g., energy harvesting \cite{ulukus2015energy},
that, in each time step, the transmitter has the option to transmit a symbol in $\Sigma_q$ (i.e.,  a protected symbol) or to refrain from communicating (represented by the $\bx$ symbol). 
When transmitting a symbol in $\Sigma_q$, the receiver will receive the transmitted symbol unless it is corrupted by an error. Otherwise, when nothing is transmitted, the symbol received may be arbitrary (due to, e.g., background noise); this uncertainty at the decoder is not counted in the error budget.
 Using this perspective, a box code of distance $d$ allows communication in the presence of any $e \leq \left\lfloor\frac{d-1}{2}\right\rfloor$ errors. Communication is terminated after $\eta$ time steps.

In a traditional communication setting, it is natural to consider the length of a codeword, i.e., the number of transmitted symbols,  as its \emph{cost} in the process of communication. In the wireless setting mentioned above, transmitting a symbol costs energy, whereas refraining from transmitting a symbol does not. With this perspective in mind, in the context of  box codes, for a codeword $c\in\cC^\bx$, the number of protected entries $\norm{c}_\bx$ corresponds to its communication cost, and the length of the code $\cC^\bx$ corresponds to the average cost of transmitting a codeword. Thus, in this context, the generalized nature of box code may imply cost benefits over traditional codes.
The extent of such benefits is the major question addressed in this work:
\begin{question}
\label{q:main}
Are there values of $M$, $d$, and $q$ for which $n^\bx_q(M,d) < n_q(M,d)$? If so, what is the maximum obtainable gap between $n_q(M,d)$ and $n_q^\bx(M,d)$?
\end{question}

In what follows, we outline the geometric interpretation of box codes, discuss prior work related to box codes on the notion of bipartite coverings of graphs, and briefly outline the results presented in this work.

\subsection{The geometric interpretation of box codes}

Box codes have interesting geometric connections to concepts studied in the context of traditional error-correcting codes, despite the fact that \( d^\bx \) neither induces a metric nor a pseudometric. This is because distinct words may have a distance of \( 0 \), and the triangle inequality does not hold.

\begin{figure}[t]
	\vspace{-10mm}
 \includegraphics[scale=0.6]{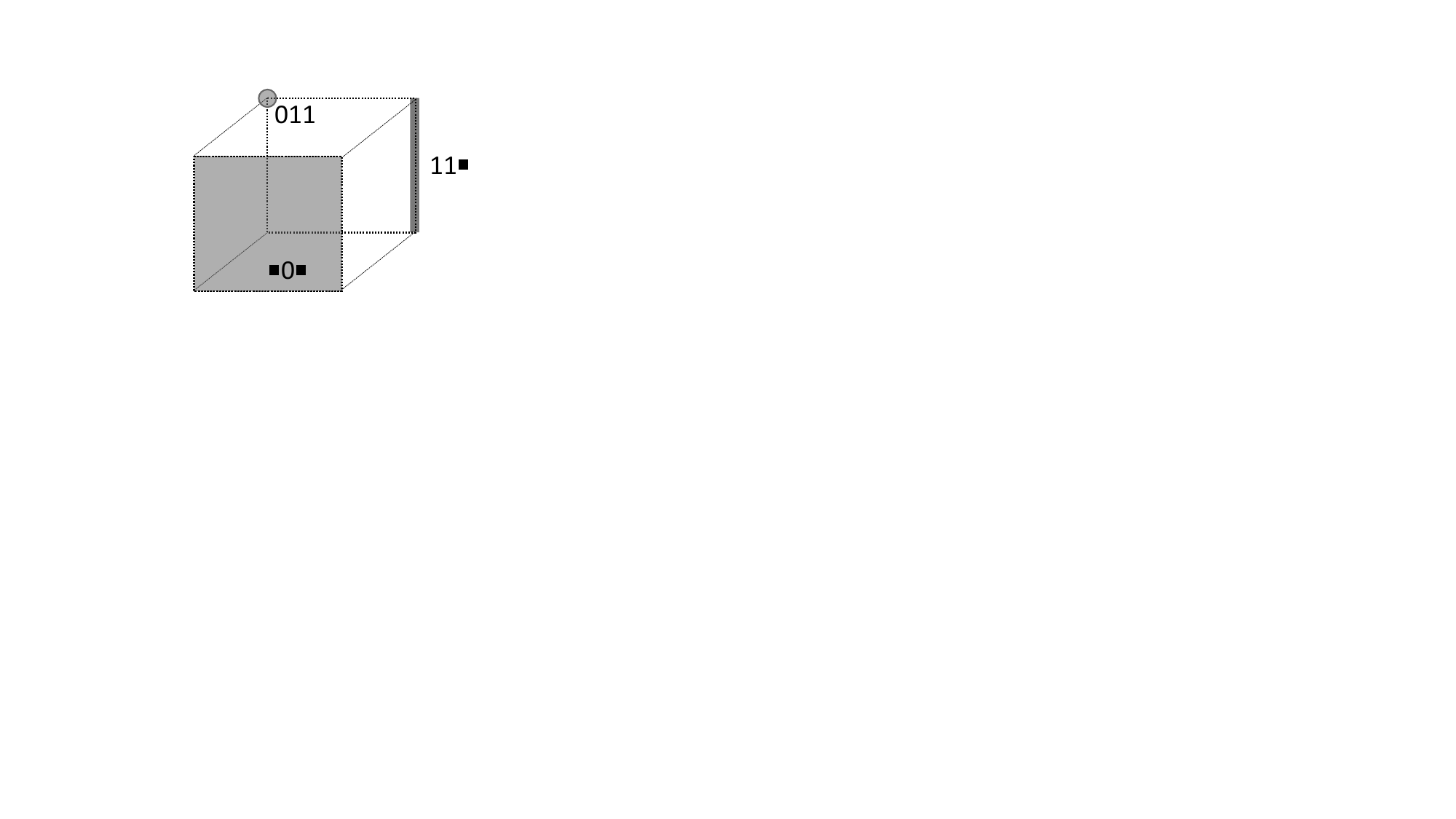}
	\vspace{-70mm}
	\caption{An example geometric representation of a $(2,3,1)_2$ box code whose protected symbols are confined to the first $\eta=3$ locations. The code restricted to $\Sigma^3$ includes three codewords $\{(0,1,1),(1,1,\bx),(\bx,0,\bx)\}$. $X((0,1,1))=(0,1,1)$ is represented by a gray point, $X((1,1,\bx))$ is represented by a gray line, and $X((\bx,0,\bx))$ is represented by a gray box of dimension 2.}
 	\label{fig:cube}
\end{figure}

Given a word \( w \in \Sigma^\eta \), we say it is \emph{consistent} with another word \( v \in \Sigma_q^\eta \) (where all entries of $v$ are protected), 
if $v_i=w_i$ whenever $w_i \ne \bx$ (i.e., $v$ and $w$ agree on all protected entries and thus $d^\bx(w, v) = 0$).
The set of all consistent words with \( w \) is denoted
\[
X(w) \eqdef \set*{v \in \Sigma_q^\eta ~:~ d^\bx(w, v) = 0}.
\]
We note that \( X(w) \) is a box of dimension \( \eta - \norm{w}_\bx \) in $\Sigma_q^\eta$.
See Figure~\ref{fig:cube} for an example binary box code.

A fundamental object in coding theory is a ball. In the metric space \( \Sigma_q^\eta \) with the Hamming distance \( d(\cdot, \cdot) \), a ball of radius \( r \) (and dimension \( \eta \)) centered at \( v \in \Sigma_q^\eta \) is defined as
\[
B_r(v) \eqdef \set*{v' \in \Sigma_q^\eta ~:~ d(v, v') \leq r}.
\]
An \( (\eta, M, d) \) code is simply a packing of \( \Sigma_q^\eta \) by balls of radius \( \floor{\frac{d-1}{2}} \) centered at the codewords.

Similarly, we can define for any word \( w \in \Sigma^\eta \) the \emph{protected ball} of radius \( r \) centered at \( w \) as
\[
B^\bx_r(w) \eqdef \set*{v' \in \Sigma_q^\eta ~:~ d^\bx(w, v') \leq r}.
\]
This is the set of all words containing only protected symbols that agree with the protected symbols of \( w \), except in at most \( r \) locations. Hence, an \( (n, M, d) \) box code, with all its protected symbols confined to the first \( \eta \) locations, is a packing of \( \Sigma_q^\eta \) by protected balls of radius \( \floor{\frac{d-1}{2}} \) centered at the codewords.

It is not difficult to see that a protected ball centered at \( w \) is the union of balls centered at words $v$
consistent with \( w \), i.e.,
\[
B^\bx_r(w) = \bigcup_{v \in X(w)} B_r(v).
\]
Thus, \( B^\bx_r(w) \) is the Cartesian product of a box of dimension \( \eta - \norm{w}_\bx \) and a ball of radius \( r \) and dimension \( \norm{w}_\bx \).

The Cartesian product of balls and boxes has important applications in coding theory. It was shown in~\cite{AhlKha98} that these shapes are the optimal anticodes in Hamming spaces, where the exact values of their dimensions is determined by $\eta$, $d$, and $q$. In particular, tiling the Hamming space with these shapes forms a diameter-perfect code~\cite{AhlAydKha01}, which is a generalization of perfect codes.
In this context, it is interesting to study packings of protected balls in the context of box codes. For box codes, when such a packing is a tiling, we say that the box code is perfect.
To better understand the geometry of box codes, we extend our study beyond Question~\ref{q:main} and discuss various constructions of perfect box codes in Section~\ref{sec:perfect} of this work.

\subsection{Connection to bipartite graph covering and prior results}
For positive integers $M$ and $d$, a bipartite $d$-covering of the complete graph $K_M$ on $M$ vertices is a (potentially infinite) collection $\cH = \set{H_1,H_2,\dots}$  of (complete) bipartite graphs, where graph $H_i$ has vertex set $V(H_i) \subseteq [M]\eqdef\set{1,\dots,M}$, such that each edge of $K_M$ appears in at least $d$ of the bipartite graphs in $\cH$. The capacity of a covering $\cH$ is defined to be 
\[\capacity_{M,d}(\cH)\eqdef \sum_i \abs*{V(H_i)}.\]
The covering capacity is then defined as 
\[\capacity(M,d)\eqdef \min_{\cH}\capacity_{M,d}(\cH),\]
and has seen prior studies in, e.g., \cite{KimLee23,Alo23,Han64}.
We show below that
\begin{equation}
\label{eq:cap}
\frac{\capacity(M,d)}{M} \eqdef \ocapacity(M,d) =  n_2^\bx(M,d).
\end{equation}

We start by presenting a bijection between $(n,M,d)_2$ box codes and bipartite $d$-coverings $\cH$ of $K_M$. Let $\cC^\bx$ be a $(n,M,d)_2$ box code. We first identify the $M$ codewords of $\cC^\bx$ with the $M$ vertices of $K_M$. We define a corresponding family $\cH=\cH(\cC^\bx)$.
For every $i$, let $H_i=(A_i,B_i)$ be the complete bipartite graph defined by 
\begin{align*}
A_i&\eqdef \set*{ c\in\cC^\bx ~:~ c_i=0}, &
B_i&\eqdef \set*{ c\in\cC^\bx ~:~ c_i=1}.
\end{align*}
We then have that each edge $\set{c,c'}\in K_M$ appears in the following number of bipartite graphs:
\begin{align*}
\abs*{\set*{i ~:~ \set{c,c'}\in H_i}} & = \abs*{\set*{i ~:~ \set{c_i,c'_i}=\set{0,1}}} \\
& = d^\bx(c,c') \geq d,
\end{align*}
where the last inequality is due to the minimum distance of $\cC^\bx$. This implies that $\cH$ is a $d$-covering $K_M$.
Moreover, it holds for the length $n$ of $\cC^\bx$ that
\[
n=\frac{1}{M}\sum_{c\in\cC^\bx} \norm{c}_\bx = \frac{\capacity_{M,d}(\cH)}{M}.
\]
It is not hard to verify that the mapping above is a bijection, implying that~\eqref{eq:cap} holds.\footnote{Formally, the bijection presented is between equivalent classes of box codes $\cC^\bx$ and those of bipartite $d$-coverings $\cH$. Here, any two codes are equivalent if the binary values in any coordinate are permuted; similarly the complete hypergraph $H_i(A_i,B_i)$ is equivalent to $H_i(B_i,A_i)$.}

Bipartite graph coverings (without explicit reference to \( n^\bx_2(M,d) \)) have seen a number of prior studies that address 
upper and lower bounds on \( \capacity(M,d) \). In particular, \cite{Han64} demonstrates that \( \ocapacity(M, d=1) \geq \log_2 M \), which implies the bound
\[
n_2^\bx(M,1) \geq \log_2 M.
\]
An analogous bound for traditional codes is trivial, as it implies that the length of any code with distance \( d=1 \) and \( M \) codewords is at least \( \log_2 M \).
\cite[Theorem 1.3]{KimLee23} establishes a lower bound, which becomes 
\begin{multline*}
n^\bx_2(M,d) \geq
\max\bigg\{2d\parenv*{1-\frac{1}{M}}, \\
\log_2{M} + \floor*{\frac{d-1}{2}} \log_2\parenv*{\frac{\log_2 M}{d}}-d-1\bigg\},
\end{multline*}
which is recently improved in \cite[Theorem 1.1]{Alo23} to
\begin{multline}
\label{eq:alon1}
n^\bx_2(M,d) \geq
\max\bigg\{2d\parenv*{1-\frac{1}{M}}, \\
\log_2{M} + \floor*{\frac{d-1}{2}} \log_2\parenv*{\frac{2\log_2 M}{d-1}}\bigg\},
\end{multline}

In \cite[Theorem 1.2]{Alo23} it was also observed that $\ocapacity(M,d) \leq n_2(M,d)$ which implies, as stated previously, that any upper bound on $n_2(M,d)$ is also one on $n_2^\bx(M,d)$. 

It is natural to generalize the above definitions to arbitrary graphs and to consider coverings of not only complete graphs. Indeed, this was also done in the aforementioned works, where a generalized notion of covering a graph \( G \) by complete bipartite graphs was introduced. Here, one designs a collection \( \cH = \{H_1, H_2, \dots\} \) of complete bipartite graphs such that each edge of \( G \) appears in at least \( d \) of the bipartite graphs in \( \cH \). 

Similarly, the generalized capacity of a covering \( \cH \) and the covering capacity of \( G \) are defined as
\[
\capacity_{G,d}(\cH) \eqdef \sum_i \abs{V(H_i)} \quad \text{and} \quad \capacity(G,d) \eqdef \min_{\cH} \capacity_{G,d}(\cH).
\]

It is not hard to verify that \( \overline{\capacity}(G,d)\triangleq \frac{\capacity(G,d)}{|G|} \) for a graph of size $M$ corresponds to the generalized notion of \( n^\bx_{G,q}(M,d) \) (or \( n_{G,q}(M,d) \)), in which one seeks the minimum-length box code \( \cC^\bx \) (or traditional code \( \cC \)) with \( M \) codewords, such that for each edge \( \set{c, c'} \in G \), the corresponding codewords satisfy \( d^\bx(c, c') \geq d \) (or \( d(c, c') \geq d \)).
Specifically, for $G$ of size $M$,
\[\ocapacity(G,d)= n^\bx_{G,2}(M,d) \leq n_{G,2}(M,d).\]
In \cite[Theorem 3]{KatSze67} it is shown that for $d=1$,
\[
\capacity(G,1) \geq \sum_{m=1}^{|G|}\log_2\left(\frac{|G|}{|G|-\Delta_m}\right),
\]
where $\Delta_m$ is the degree of vertex $m$ in $G$, and in \cite{Alo23} it is shown that 
\begin{align}
\label{eq:alon2}
\capacity(G,1) \geq \sum_{m=1}^{|G|}\log_2\left(\frac{|G|}{\alpha_i}\right),
\end{align}
where $\alpha_i$ is the maximum size of an independent set in $G$ that contains $i$.
Following \cite{Alo23}, we note that $\alpha_i \leq |G|-\Delta_i$, so the latter bound improves on the former.

\subsection{Overview}
In Section~\ref{sec:gap}, we address Question~\ref{q:main} and present a number of box code constructions for which $n^\bx_q(M,d)$ is strictly smaller than $n_q(M,d)$ (for various values of $q$, $M$, and $d$).
In Section~\ref{sec:perfect}, we extend the constructions of \cite{AhlAydKha01}
and present a number of perfect box-code constructions.
Finally, in Section~\ref{sec:gen}, we present new bounds on $n^\bx_{G,q}(M,d)$, extending prior results of \cite{Alo23}.

\section{A Gap Between $n_q(M,d)$ and $n^\bx_q(M,d)$}
\label{sec:gap}

The goal of this section is to provide evidence for a gap, as asked in Question~\ref{q:main}. Specifically, we show a gap of $1-o(1)$ between $n_q(M,d)$ and $n^\bx_q(M,d)$ in some asymptotic regimes, by designing box codes whose traditional counterparts must be longer. Our strategy for the design of box codes is to start with a traditional code with good parameters, remove some of its codewords, and insert $\bx$'s in some positions.

The first construction relies on binary Hamming codes, and the fact that they attain the ball-packing bound with equality.

\begin{construction}
\label{con:ham}
Let $\cC$ be the $[2^m-1,2^m-m-1,3]$ linear binary Hamming code, and denote 
\begin{align}
\label{eq:hamnM}
n&\eqdef 2^m-1, & M& \eqdef \abs*{\cC}=2^{2^m-m-1}=\frac{2^n}{n+1}.
\end{align}
Construct the box code $\cC^\bx$ to contain all the codewords of $\cC$ that have even weight, as well as $\ceil{\frac{1}{2n}M}$ arbitrary codewords of the odd-weight  $\frac{n-1}{2}=2^{m-1}-1$. Then, replace the first coordinate with $\bx$ in all the codewords of (even) 
weight $w$ that satisfy
\[ \abs*{ w-\frac{n-1}{2}} \geq 4.\]
Finally, append $\bx^\infty$ to all codewords.
\end{construction}

\begin{lemma}
\label{lem:ham}
The box code $\cC^\bx$ from Construction~\ref{con:ham} is well defined, for all sufficiently large $m$, and has parameters $(n^\bx,M^\bx,d^\bx)_2$ with
\begin{equation}
\label{eq:hamparm}
\begin{split}
n^\bx &\leq n-1+O(1/\sqrt{\log M^\bx}),\\
M^\bx &= \frac{1}{2}M+\ceil*{\frac{1}{2n}M},\\
d^\bx &\geq 3,
\end{split}
\end{equation}
where $n$ and $M$ are given in~\eqref{eq:hamnM}.
\end{lemma}
\begin{IEEEproof}
We first make sure the code is well defined. The only possible point of difficulty is the existence of sufficiently many  codewords of weight $\frac{n-1}{2}$ in $\cC$. Denote by $A_i$ the number of codewords of weight $i$ in $\cC$. According to~\cite[Proposition 4.1]{EtzVar94},
\begin{equation}
\label{eq:hamenum}
\begin{split}
    A_i & = \frac{\binom{n}{i}+n\Delta_i}{n+1},\\
    \Delta_i & = \begin{cases}
        \binom{\frac{n-1}{2}}{\floor{i/2}} & i\equiv 0,3\pmod{4}, \\
        -\binom{\frac{n-1}{2}}{\floor{i/2}} & i\equiv 1,2\pmod{4}.
    \end{cases}
\end{split}
\end{equation}
In our case, for $m\geq 3$, we have $i=\frac{n-1}{2}\equiv 3\pmod{4}$. Hence,
\begin{align*}
    A_{\frac{n-1}{2}} & = \frac{\binom{n}{\frac{n-1}{2}}+n\binom{\frac{n-1}{2}}{\floor{(n-1)/4}}}{n+1} \\
    & \geq \frac{1}{n+1}\binom{n}{\frac{n-1}{2}}
    =\frac{1}{n+1}\cdot\frac{2^n}{\sqrt{\pi n/2}}(1-o(1)) \\
    & > \frac{2^n}{2n(n+1)} = \frac{1}{2n}M,
\end{align*}
for all sufficiently large $m$. We used here a standard  approximation for the central binomial coefficient, as well as the fact that $M=\frac{2^n}{n+1}$. Thus, we have sufficiently many odd-weight codewords of weight $\frac{n-1}{2}$ in $\cC$ to choose from.

The cardinality of the code is immediate from the construction, since $M/2$ of the codewords of the original code have even weight. Next, we examine the distance $d^\bx$. The (standard) distance between two codewords of odd weight in $\cC$ is at least $4$, since it must be even, and in addition $\cC$ has minimum distance $3$. A similar statement can be made for the distance between two codewords of even weight in $\cC$.
Thus, the corresponding codewords in $\cC^\bx$ have $d^\bx$-distance at least $3$. This follows from the fact that for some codewords in $\cC^\bx$ the first coordinate has been replaced by $\bx$.
Finally, assume $x,y\in \cC$ are codewords with differing parity, with $x$ having odd weight $\frac{n-1}{2}$, and $y$ having even weight $w$. If $\abs{w-(n-1)/2}\leq 3$, then no $\bx$ is present in $y$, and $d^\bx(x,y)\geq 3$ by virtue of the minimal distance of $C$. Otherwise, $d(x,y) \geq \abs{w-(n-1)/2}\geq 4$, and necessarily the corresponding codewords in $\cC^\bx$ have $d^\bx$ distance at least 3 (given the presence of $\bx$ in the first coordinate of the codeword in $\cC^\bx$ corresponding to $y$).

Finally, we bound the length $n^\bx$ of the box code $\cC^\bx$. We note that all codewords of $\cC^\bx$ contain a protected bit in coordinates $2$ to $n$. Additionally, the first coordinate holds a protected bit in the codewords originating from odd-weight codewords in $\cC$, and those with even weight $\frac{n-1}{2}+\set{1,-1,3,-3}$. Thus,
\begin{equation}
\label{eq:nbox}
n^\bx = n-1+\frac{1}{M^\bx}\parenv*{\ceil*{\frac{1}{2n}M}+\sum_{i\in\set{1,-1,3,-3}}A_{\frac{n-1}{2}+i}}.
\end{equation}
Since the all-ones vector is a codeword of $\cC$, we have a symmetry in the weight enumerator,
\[
A_i = A_{n-i}.
\]
Thus,
\[
\sum_{i\in\set{1,-1,3,-3}}A_{\frac{n-1}{2}+i} = \sum_{i=0}^3A_{\frac{n-1}{2}-i}.
\]
Additionally, for $m\geq 5$ the weight enumerator of the binary Hamming code is log-concave~\cite[Theorem 3]{ShiWanAnKim24}, and therefore unimodal. Thus,
\[
\sum_{i\in\set{1,-1,3,-3}}A_{\frac{n-1}{2}+i} = \sum_{i=0}^3A_{\frac{n-1}{2}-i} \leq 4 A_{\frac{n-1}{2}}.
\]
Once again, by~\eqref{eq:hamenum} and the standard approximation for the central binomial coefficient,
\begin{align*}
    A_{\frac{n-1}{2}} & = \frac{\binom{n}{\frac{n-1}{2}}+n\binom{\frac{n-1}{2}}{\floor{(n-1)/4}}}{n+1} \\
    & = \parenv*{\frac{1}{n+1}\cdot\frac{2^n}{\sqrt{\pi n/2}}+\frac{n}{n+1}\cdot\frac{2^{\frac{n-1}{2}}}{\sqrt{\pi n/4}}}(1+o(1)) \\
    & \leq \frac{1}{n+1}\cdot\frac{2^n}{\sqrt{\pi n/2}}(1+o(1)) \\
    & \leq \frac{M}{\sqrt{n}}(1+o(1)).
\end{align*}
Plugging this back into~\eqref{eq:nbox} and noting that $M=\Theta(M^\bx)$, we have
\[
n^\bx \leq n-1+O(1/\sqrt{n}),
\]
which proves our claim, since $\log M^\bx = \Theta(\log M) = O(n)$.
\end{IEEEproof}

\begin{theorem}
For all sufficiently large values of $M^\bx$ as in~\eqref{eq:hamparm}, we have
\[
n^\bx_2(M^\bx,3) \leq n_2(M^\bx,3) - 1 + O(1/\sqrt{\log M^\bx}).
\]
\end{theorem}
\begin{IEEEproof}
Consider the $(n^\bx,M^\bx,d^\bx)_2$ box code from Construction~\ref{con:ham}. By Lemma~\ref{lem:ham} we have
\[
n^\bx_2(M^\bx,3) \leq n^\bx \leq n-1+O(1/\sqrt{\log M^\bx}).
\]
We now contend that a traditional binary code of size $M^\bx$ and minimum distance at least $3$ must have length at least $n$, i.e.,
\[ n \leq n_2(M^\bx,3).\]
Assume to the contrary that is not the case and a code with this cardinality and minimum distance is possible with length $n-1$. In this space, the volume of a ball of radius $1$ is $n$. But then,
\begin{align*}
M^\bx n &= \parenv*{\frac{1}{2}M+\ceil*{\frac{1}{2n}M}}n \\
&> \frac{n2^{n-1}}{n+1}+\frac{n2^{n-1}}{n(n+1)} = 2^{n-1},
\end{align*}
where we used the fact that, for $m\geq 2$, $M$ is a power of $2$ while $n$ is not, so $n\nmid M$. However this contradicts the ball-packing bound~\cite[Ch.~1, Theorem 6]{MacSlo78} implying that $n(M^\bx,3)> n-1$ or, equivalently, $n(M^\bx,3) \geq n$.
Thus,
\begin{align*}
n^\bx_2(M^\bx,3) &\leq n^\bx \leq n-1+O(1/\sqrt{\log M^\bx}) \\
&\leq n_2(M^\bx,3)-1+O(1/\sqrt{\log M^\bx}),
\end{align*}
which completes the proof.
\end{IEEEproof}

A similar approach to that of Construction~\ref{con:ham} may be applied to Reed-Solomon (RS) codes. In this instance, we shall be using that fact that RS codes attain the Singleton bound with equality.

\begin{construction}
\label{con:rs}
Let $\cC_1$ be an $[n,k,d]_q$ RS code, where $k\geq 2$, $n=k+d-1$, and $q\geq n+1$ is a prime power. Additionally, let $\cC_2\subseteq\cC_1$ be an $[n,k-1,d+1]_q$ RS code.

Construct the box code $\cC^\bx$ in the following way.
First let
\[
\cC^\bx = \cC_2 \cup \set*{c^*},
\]
where $c^*\in \cC_1\setminus \cC_2$ is arbitrary. Define
\[
\cC'_2 \eqdef \set*{c\in\cC_2 ~:~ d(c,c^*) = d}.
\]
Then, set $\bx$ as the first coordinate of all the codewords of $\cC^\bx$ from $\cC_2\setminus \cC'_2$. Finally, append $\bx^\infty$ to all codewords.
\end{construction}

\begin{lemma}
\label{lem:rs}
The box code $\cC^\bx$ from Construction~\ref{con:rs} has parameters $(n^\bx,M^\bx,d^\bx)_q$ with
\begin{equation}
\label{eq:rsparm}
\begin{split}
n^\bx & \leq k+d-2 + \frac{1+(q-1)\binom{k+d-1}{d}}{q^{k-1}+1}, \\
M^\bx & = q^{k-1}+1,\\
d^\bx & \geq d,
\end{split}
\end{equation}
where $k$, $d$, and $q$, satisfy the requirements of Construction~\ref{con:rs}.
\end{lemma}

\begin{IEEEproof}
By construction, it follows that $M^\bx=q^{k-1}+1$. For the distance $d^\bx$, we first note that the minimum distance between codewords in $\cC_2$ is $d+1$, and thus in $\cC^\bx$ the $d^\bx$-distance between pairs of codewords corresponding to codewords in $\cC_2$ is at least $d$.
The codeword $c^*$ is at distance at least $d$ from any of the codewords in $\cC_2$, because they all are part of $\cC_1$ which has minimum distance $d$. By construction, those codewords of $\cC_2$ that are at distance $d$ from $c^*$, do not get changed, and so the minimum distance of $\cC^\bx$ is at least $d$.

Finally, we bound the length $n^\bx$. We note that all $M^\bx$ codewords of $\cC^\bx$ have protected symbols in positions $2$ to $n=k+d-1$. In the first position, exactly $1+\abs{\cC'_2}$ codewords have a protected symbol. 
We upper bound $\abs{\cC'_2}$ by the number of codewords of $\cC_1$ that are at distance $d$ from $c^*$. By the linearity of the code $\cC_1$, the latter number equals the number of codewords of $\cC_1$ of minimal weight, i.e., weight $d$. The number of minimal weight codewords in any $[n,k,d]$ MDS code is known to be $(q-1)\binom{n}{d}$ (e.g., see~\cite[Ch.~11.3, Corollary 5]{MacSlo78}). Thus, the number of protected symbols in the first coordinate is
\[
1+\abs*{\cC'_2} \leq 1+(q-1)\binom{k+d-1}{d}.
\]
Plugging these facts in the definition of $n^\bx$ we obtain the desired result.
\end{IEEEproof}

\begin{theorem}
\label{th:gaprs}
Let $k$, $d$, and $q$, be as in Construction~\ref{con:rs}. Then
\begin{multline*}
n^\bx_q(q^{k-1}+1,d) \\
\leq n_q(q^{k-1}+1,d)-1 + \frac{1+(q-1)\binom{k+d-1}{d}}{q^{k-1}+1}.
\end{multline*}
\end{theorem}
\begin{IEEEproof}
The Singleton bound~\cite[Ch.~11, Problem (3)]{MacSlo78} implies
\[
n_q(q^{k-1}+1,d) \geq \ceil*{\log_q(q^{k-1}+1)} + d -1 \geq k+d-1.
\]
Combining this with~\eqref{eq:rsparm} gives the desired result.
\end{IEEEproof}

We mention that in all asymptotic regimes pertinent to Theorem~\ref{th:gaprs}, we have $q\to\infty$, since $q\geq d+k$. In some of the regimes, the rightmost addend in the claim of Theorem~\ref{th:gaprs} is $o(1)$. In particular, $n^\bx_q(q^{k-1}+1,d) \leq n_q(q^{k-1}+1,d)-1+o(1)$ when:
\begin{itemize}
\item
$d$ is constant, and $k,q\to\infty$.
\item
$k\geq 3$ is constant, $d,q\to\infty$, and $d=o(q^{\frac{k-2}{k-1}})$.
\item
$d,k,q\to\infty$ and $d=\Theta(k)$.
\end{itemize}

We finally note that a simple construction shows a gap between $n_2(M,d)$ and $n^\bx_2(M,d)$ for $d=1$ as well. 
Consider the trivial $(n,2^n,1)_2$ code $\cC$ consisting of all codewords in $\set{0,1}^n$.
Define $\cC^\bx$ by first removing all even-weight codewords in $\cC$, except for the all-zero codeword. Then, we change the first coordinate of all codewords of weight at least $3$ to $\bx$, and finally we append $\bx^\infty$ to all codewords. The resulting box code $\cC^\bx$ has parameters  $(n^\bx,2^{n-1}+1,1)_2$ as follows. First, the number of codewords equals $2^{n-1}+1$ by construction. As for the minimum distance of $\cC^\bx$, this follows from the fact that any two codewords of odd weight from $\cC$ have Hamming distance at least $2$ and their corresponding codewords in $\cC^\bx$ thus have $d^\bx$-distance at least $1$. Moreover, all codewords originating from odd-weight vectors have at least one surviving entry $1$, and so the all-zero codeword is at distance at least $1$ from all other codewords of $\cC^\bx$. Finally, by simple calculation,
\[
n^\bx = n-1+\frac{n+1}{2^{n-1}+1} = n-1+O(n/2^n).
\]
Thus
\begin{align*}
n^\bx_2(2^{n-1}+1,1)_2 
&\leq n-1+O(n/2^n) \\
&\leq n_2(2^{n-1}+1,1)_2-1+O(n/2^n),
\end{align*}
where, in the second inequality above, we use the fact  that $n_2(M,1)_2 = \ceil{\log_2(M)}$.
Note that the construction of $\cC^\bx$ above is close to optimal as \cite{Han64} shows that $n^\bx_2(M,1)_2 \geq \log_2{M}$, which for $M=2^{n-1}+1$ gives $n^\bx_2(M,1) \geq n-1+O(1/2^n)$.

\section{Perfect Codes}
\label{sec:perfect}

As already mentioned in the introduction, the error-balls of box codes are, in fact, a Cartesian product of a ball and a box, the dimensions of which are determined by the number of $\bx$ elements in the codeword around which the ball is centered. Unlike traditional coding, in which the Hamming metric determines that all balls of a given radius have the same size, here balls may differ in size due to differing numbers of $\bx$ elements in their centers. Even though we have this added complication, some non-degenerate perfect codes 
may still be constructed.

We first construct perfect box codes of minimum distance $1$. To present our construction we need the following notation. Let $v\in\Sigma^n$ be some vector, and let $w\in\Sigma^\infty$ be an infinite sequence. Then $vw$ denotes the concatenation of $v$ and $w$. Naturally, if $\cC^\bx$ is a box code over $\Sigma$, then we define
\[
v\cC^\bx \eqdef \set*{ vc ~:~ c\in\cC^\bx}.
\]

\begin{construction}
\label{con:perfect}
Assume that for every letter $a\in\Sigma_q$, we fix some perfect box code $\cC^\bx_a\subseteq\Sigma^\infty$ with parameters $(n_a,M_a,1)_q$ and protected length $\eta_a$. We construct the following box code,
\[
\cC^\bx = \bigcup_{a\in\Sigma_q} a\cC^\bx_a.
\]
\end{construction}

\begin{theorem}
The code $\cC^\bx$ from Construction~\ref{con:perfect} is a perfect box code with parameters $(n,M,1)_q$ and protected length $\eta$, where
\begin{align*}
n &= \frac{\sum_{a\in\Sigma_q} n_a M_a}{\sum_{a\in\Sigma_q} M_a},\\
M &= \sum_{a\in\Sigma_q} M_a,\\
\eta &= 1 + \max_{a\in\Sigma_q}\eta_a.
\end{align*}
\end{theorem}

\begin{IEEEproof}
The size of the code, $M$, is immediate by construction. Since the new first coordinate of the code contains only protected symbols, the expression for $\eta$ also follows. We note that the total number of protected symbols in each $\cC_a$ is exactly $n_a M_a$, and thus, after counting all protected symbols in all component codes and dividing by the number of codewords, $M$, we obtain the length of the code, $n$.

Finally, we show that $\cC$ is a perfect code. For any arbitrary vector of protected symbols, $v=(v_1,v_2,\dots,v_\eta)\in\Sigma_q^\eta$, we show that there exists a unique codeword $c\in\cC$ such that $v\bx^\infty\in B^\bx_0(c)$. Given the first coordinate, $v_1$, it suffices to check if there is a unique $c'\in\cC_{v_1}$ such that $v\bx^\infty\in B^\bx_0(v_1 c')$. The last claim is guaranteed by the fact that $\cC_a$ is an $(n_a,M_a,1)_q$ perfect box code.
\end{IEEEproof}

\begin{example}
Let $\Sigma_q=\set{0,1}$. We now use Construction~\ref{con:perfect} iteratively, starting with $\cC^\bx(0)=\set{\bx^\infty}$, and at iteration $i=1,2,\dots$, we pick $\cC^\bx_0=\cC^\bx(i-1)$ and $\cC^\bx_1=\set{\bx^\infty}$, to obtain $\cC^\bx(i)$. After $n$ steps, it follows that
\[
\cC^\bx(n) = 
\set*{
\begin{matrix}
000\dots000,\\
000\dots001,\\
000\dots01\bx,\\
000\dots1\bx\bx,\\
\vdots\\
01\bx\dots\bx\bx\bx,\\
1\bx\bx\dots\bx\bx\bx\phantom{,}
\end{matrix}
}\bx^\infty
\]
is a perfect $(\frac{n}{2}+1-\frac{1}{n},n+1,1)_2$ box code.
\end{example}

The last example is clearly a non-degenerate box code, since it is not merely a traditional code concatenated with $\bx^\infty$. It is also interesting in that it provides a tiling of the space with boxes of different dimensions.

Next, we present a construction for non-degenerate binary perfect box codes with distance $3$. This construction makes use of binary nearly-perfect $1$-covering codes~\cite{BorEtzRot24}.

Set $\Sigma_2=\F_2=\set{0,1}$. A (traditional) $(n,M)_2$ code, $\cC$, is said to have covering radius $R$ if
\[
\bigcup_{c\in\cC} B_R(c) = \Sigma_2^n.
\]
The simple ball-covering bound states that
\[
M\sum_{i=0}^R \binom{n}{i} \geq 2^n,
\]
and codes attaining it with equality are said to be perfect. This bound was improved by Van Wee~\cite{Wee88} to 
\begin{equation}
\label{eq:wee}
M\cdot\parenv*{
\sum_{i=0}^R\binom{n}{i}-\frac{\binom{n}{R}}{\ceil*{\frac{n-R}{R+1}}}\parenv*{\ceil*{\frac{n+1}{R+1}}+\frac{n+1}{R+1}}}\geq 2^n.
\end{equation}
When $R+1|n+1$, \eqref{eq:wee} reduces to the ball-covering bound. When $R+1\nmid n+1$, a code attaining~\eqref{eq:wee} with equality is said to be a \emph{nearly perfect $R$-covering code}. Specifically, for $R=1$ and even $n$, a nearly perfect $1$-covering code (NP1CC) satisfies
\begin{equation}
\label{eq:wee1}
M = \frac{2^n}{n}.
\end{equation}

In the terminology of~\cite{BorEtzRot24}, a binary NP1CC of type A, $\cC$, has the following added property: its codewords may be partitioned uniquely into disjoint pairs, $\set{c,c'}$, such that $d(c,c')=1$. We call the set of these pairs the \emph{canonical partition} of $\cC$. We use such codes for our next construction.

A notation we need for the construction is the following: Assume we have two binary vectors that differ only in the $i$th position,
\begin{align*}
c&=(c_1,\dots,c_{i-1},0,c_{i+1},\dots,c_n),\\
c'&=(c_1,\dots,c_{i-1},1,c_{i+1},\dots,c_n).
\end{align*}
We define the \emph{meshing} of $c$ and $c'$ as
\[
c'\boxdot c = c \boxdot c' \eqdef (c_1,\dots,c_{i-1},\bx,c_{i+1},\dots,c_n).
\]

\begin{construction}
\label{con:np1cc}
Let $\cC$ be an $(n,\frac{2^n}{n})_2$ binary NP1CC of type A, $n$ even, and let $\cP$ be its canonical partition. We construct the following box code
\[
\cC^\bx = \set*{c\boxdot c' ~:~ \set{c,c'}\in\cP}\bx^\infty.
\]
\end{construction}

\begin{theorem}
The code $\cC^\bx$ from Construction~\ref{con:np1cc} is a binary perfect $(n-1,\frac{2^{n-1}}{n},3)_2$ box code.
\end{theorem}
\begin{IEEEproof}
Since the NP1CC contains $\frac{2^n}{n}$ codewords, its canonical partition contains $\frac{2^{n-1}}{n}$ pairs, each becoming a codeword in $\cC^\bx$, hence the size of $\cC^\bx$. Each of the codewords of $\cC^\bx$ contains a single $\bx$ entry, giving us the length $n-1$.

Finally, we show that $\cC^\bx$ is perfect with minimum distance $3$. We do so by showing that the protected balls $B^\bx_1(\cdot)$, centered at the codewords of $\cC^\bx$, tile $\Sigma_2^n$. Let $v\in\Sigma_2^n$ be some vector. Since $\cC$ is a $1$-covering code, there exists $c\in\cC$ such that $d(c,v)\leq 1$. Let $c'\in\cC$ be such that $\set{c,c'}\in\cP$. By construction, the codeword $(c\boxdot c')\bx^\infty\in\cC^\bx$ satisfies,
$d^\bx(c\boxdot c',v)\leq 1$, and so
\[
v\in B^\bx_1(c\boxdot c').
\]
Since each codeword $c''\in\cC^\bx$ contains a single $\bx$, we have
\[
\abs*{B^\bx_1(c'')} = 2n.
\]
But
\[
2n\cdot \abs{\cC^\bx}=2n\cdot \frac{2^{n-1}}{n} = 2^n = \abs*{\Sigma_2^n},
\]
and therefore the protected balls of radius $1$ around the codewords of $\cC^\bx$ tile $\Sigma_2^n$ and the code is perfect.
This, in turn, implies that the minimum distance of the code is 3.
\end{IEEEproof}

We remark that in Construction~\ref{con:np1cc}, if all the pairs in the canonical partition of the NP1CC differ in the same coordinate, the resulting box code $\cC^\bx$ will have all of the $\bx$ symbols in the same coordinate, thereby rendering it degenerate. However, \cite[Section V]{BorEtzRot24} constructs \emph{balanced} NP1CCs, in which the distribution of the differing coordinate in the pairs of the canonical partition is uniform over $\set{1,\dots,n}$. In particular, \cite[Section V]{BorEtzRot24} proves the existence of balanced binary NP1CCs of length $n=2^\ell$ for all $\ell\geq 3$, giving us the following corollary:

\begin{corollary}
There exist perfect binary non-degenerate $(n-1,\frac{2^{n-1}}{n},3)_2$ box codes for all $n=2^\ell$, $\ell\geq 3$.
\end{corollary}

\begin{example}
Take $n=2^\ell=8$, and use the balanced NP1CC given in~\cite[Section V]{BorEtzRot24} together with Construction~\ref{con:np1cc}, to get the following perfect $(7,16,3)_2$ box code:
\[
\cC^\bx=
\set*{
\begin{matrix}
0001101\bx,\quad 1110010\bx, \\
001101\bx1,\quad 110010\bx0, \\
01101\bx11,\quad 10010\bx00, \\
1101\bx111,\quad 0010\bx000, \\
101\bx1110,\quad 010\bx0001, \\
01\bx11100,\quad 10\bx00011, \\
1\bx111001,\quad 0\bx000110, \\
\bx1110010,\quad \bx0001101\phantom{,}
\end{matrix}
}\bx^\infty.
\]
\end{example}

\section{Covering of General Graphs}
\label{sec:gen}

We now address lower bounds on $n^\bx_{G,2}(M,d)=\ocapacity(G,d)$ for general undirected graphs $G$. 
Theorem~\ref{th:hamming-general} below extends  bounds \eqref{eq:alon1} and \eqref{eq:alon2} to general graphs $G$ with $d \geq 2$.
Specifically, our results leverage the proof techniques of \cite[Theorems 1.1 and 1.4]{Alo23} to extend the results of \cite{Alo23}.

\begin{theorem}[Lower bound on $n^\bx_{G,2}(M,d)$]
\label{th:hamming-general}
For any graph $G$ with vertex set of size $M$
and for any distance $d$, $n^\bx_{G,2}(M,d)$ is at least 
\begin{align*}
\max\bigg\{&\frac{2d}{\alpha(G)}-\frac{2d}{M},\\
&  \ \log_2\parenv*{\frac{M}{\alpha(G)}} + \floor*{\frac{d-1}{2}} \log_2\parenv*{\frac{2\log_2(M/\alpha(G))}{d-1}}, \\
& \ \floor*{\frac{d-1}{2}}+ \frac{1}{M}\sum_{u=1}^M \log_2\parenv*{\frac{M}{\alpha_u(G)}}\bigg\},
\end{align*}
where $\alpha(G)$ denotes the independence number of $G$, and $\alpha_u(G)$ denotes the size of the largest independent set in $G$ that includes the vertex $u$. \end{theorem}

\begin{IEEEproof}  
We begin by proving the result for  graphs with no isolated vertices. Let $\cC^\bx$ be an $(n,M,d)_2$ box code corresponding to $G=(V,E)$.
Namely, every codeword $c$ of $\cC^\bx$ corresponds to a vertex $v$ of $G$ and, in addition, codewords $c_u$ and $c_v$ corresponding to an edge $(u,v) \in E$ satisfy  $d^\bx(c_u,c_v) \geq d$.
Consider a bipartite graph covering $\cH=\{H_1,\dots,H_\eta\}$ corresponding to $\cC^\bx$.
Here, $\eta$ is the protected length of $\mathcal{C}^\bx$. 
Specifically, for every $i$, $H_i=(A_i,B_i)$ where $A_i=\{c \in \cC^\bx : c_i=0\}$ and $B_i=\{c \in \cC^\bx : c_i=1\}$. 
Let $h_i = |A_i|+|B_i|$ be the number of vertices in $H_i$. The number of edges in $H_i$ equals $|A_i||B_i| \leq h_i^2/4$.
The properties of $\cC^\bx$ imply that every edge $e$ of $G$ appears in at least $d$ bipartite graphs in $\cH$; i.e., $\sum_{i=1}^{\eta}h_i^2/4 \geq d|E|$.
Moreover, the block length $n=\frac{1}{M}\sum_{c \in \cC^\bx} \|c\|_\bx$ of $\cC^\bx$ equals $\frac{1}{M}\sum_i^\eta h_i$.
We conclude that 
\[
n=\frac{1}{M}\sum_{i=1}^\eta h_i \geq \frac{1}{M^2}\sum_{i=1}^\eta h_i^2 \geq \frac{4d|E|}{M^2},
\]
where for the first inequality we use the fact that $M \geq h_i$.
Now, to obtain the first asserted lower bound, notice that by Tur\'an's theorem \cite{turan1941external}, 
$|E| \geq \frac{M^2}{2\alpha(G)}-\frac{M}{2}$.

For the second asserted bound, 
let $r = \lfloor (d-1)/2 \rfloor$ and for  $v\in \{0,1\}^\eta$
let 
\[
N_v = \{ c \in \mathcal{C}^\bx : v \in B^\bx_r(c)\}.
\]
Note that if $N_v$ contains two codewords of $\cC^\bx$ corresponding to adjacent vertices in $G$, then their $d^\bx$-distance would be less than $d$, violating the $d$-covering property.  Thus, the subset of vertices in $V$ corresponding to $N_v$ is an independent set in $G$. Since the largest independent set has size $\alpha(G)$, we have $|N_v| \le \alpha(G)$.

In addition, it holds that
\[
\sum_{c \in \mathcal{C}^\bx} |B^\bx_r(c)| = \sum_{v \in \{0,1\}^\eta} |N_v| \leq \sum_{v \in \{0,1\}^\eta} \alpha(G) = \alpha(G) 2^\eta.
\]
Moreover, the size of a protected ball $B^\bx_r(c)$ is  bounded from below as follows  
\begin{equation}
    \label{bound-on-ball}
|B^\bx_r(c)| = 2^{\eta-\|c\|_\bx} \sum_{i=0}^r \binom{\|c\|_\bx}{i} \geq 2^{\eta-\|c\|_\bx} \left(\frac{\|c\|_\bx}{r}\right)^r.
\end{equation}
By convexity applied to the function $f(x) = 2^{-x} (x/r)^r$, and the fact that $n = \frac{1}{M}\sum_c \|c\|_\bx$, we obtain
\[
\frac{1}{2^\eta M}\sum_{c \in \mathcal{C}^\bx} |B^\bx_r(c)| \geq 2^{-n}(n/r)^r .
\]
By combining the two bounds we have 
\[
\frac{\alpha(G)}{M} \geq \frac{1}{2^\eta M}\sum_{c \in \mathcal{C}^\bx} |B^\bx_r(c)| \geq 2^{-n}(n/r)^r,
\]
implying that 
%
\begin{equation}
\label{stam}
\log_2\left(\frac{\alpha(G)}{M}\right) \geq -n + r\log_2\left(\frac{n}{r}\right).
\end{equation}

Since every vertex in $G$ has at least one neighbor, for every codeword $c$ in $\cC^\bx$ there exists at least one codeword $c'\in \cC^\bx$ for which $d^\bx(c,c') \geq d$. Implying that $\|c\|_\bx \geq d^\bx(c,c') \geq d$, and, in turn, that $n = \frac{1}{M}\sum_c \|c\|_\bx \geq d$.
Therefore by \eqref{stam} $n \geq \log_2(M/\alpha(G))$ and in addition
\[
n \geq  \log_2\left(\frac{M}{\alpha(G)}\right) + \left\lfloor\frac{d-1}{2}\right\rfloor\log_2\left(\frac{2\log_2(M/\alpha(G))}{d-1}\right).
\]

For the third asserted bound, 
again, let $r = \lfloor (d-1)/2 \rfloor$.
As before, for each codeword $c\in\cC^\bx$ consider the ball $B_r^\bx(c)$ and for each 
$v\in \{0,1\}^\eta$ consider $N_v$.
Recall that $N_v$ is an independent set in $G$.
Moreover, for each $c \in N_v$, $N_v$ is an independent set that includes $c$ and is thus of size at most $\alpha_c(G)$.

We will need the following special case of a lemma from \cite{Alo23}.
\begin{lemma}\cite[Lemma 3.1]{Alo23}
\label{alon-lemma}
    Let  $\{E_c\}_{c \in \cC}$ be a collection of events in the uniform probability space over $\{0,1\}^\eta$, such that  for every point $v\in \{0,1\}^\eta$, if $v \in E_c$ then the total number of events $E_{c'}$ in the collection that contain $v$ is at most $\alpha_c$. Then 
\[
1 \geq \sum_c \frac{\Pr[E_c]}{\alpha_c}  
\]

\end{lemma}
Returning to the proof of the theorem. By Lemma \ref{alon-lemma},
where the event $E_c$ is defined to be the subset $B_r^\bx(c)$ and $\alpha_c$ is defined to be $\alpha_c(G)$,
we conclude that,
\begin{align}
1 & \geq  \sum_c \frac{\Pr[E_c]}{\alpha_c}  = \sum_c \frac{|B_r^\bx(c)|}{2^\eta \alpha_c(G)} \nonumber\\
& \geq
\sum_c 2^{-\|c\|_\bx-\log_2{\alpha_c(G)}+r\log_2 \left(\frac{\|c\|_\bx}{r}\right)}\label{stamstam}, 
\end{align}
where \eqref{stamstam} follows from \eqref{bound-on-ball}.
 The above now implies by the arithmetic-geometric means inequality that
\begin{align*}
1 & \geq 
M2^{\frac{\sum_c \left( -\|c\|_\bx-\log_2{\alpha_c(G)}+r\log_2 \left(\frac{\|c\|_\bx}{r}\right)\right)}{M}}\\
& =
2^{\frac{\sum_c \left( -\|c\|_\bx+\log_2{(M/\alpha_c(G))}+r\log_2 \left(\frac{\|c\|_\bx}{r}\right)\right)}{M}},
\end{align*}
which, to conclude the asserted bound, as $n = \frac{1}{M}\sum_c \|c\|_\bx$, implies that
\begin{align*}
n & \geq \frac{1}{M} \left(\sum_c \log_2{\left(\frac{M}{\alpha_c(G)}\right)}
+\sum_c r\log_2 \left(\frac{\|c\|_\bx}{r}\right)\right) \\
& \geq r+ \frac{1}{M} \sum_c \log_2{\left(\frac{M}{\alpha_c(G)}\right)},
\end{align*}
where in the last inequality we use the observation, as before, that, for every $c \in \cC^\bx$, $\|c\|_\bx \geq d > 2r$.

Lastly, we extend the result to arbitrary graphs. Assume that the graph $G$ contains $k$ isolated vertices, and let $G'$ be the graph on $M' = M - k$ vertices obtained by removing these $k$ isolated vertices from $G$. 

We then have the following bound
\[
n^\bx_{G,2}(M,d) \geq f\big(\alpha(G'), M', \{\alpha_u(G') : u \in G'\}\big),
\]
where \( f\big(\alpha(G'), M', \{\alpha_u(G') : u \in G'\}\big) \) is the lower bound established for graphs without isolated vertices, applied to the graph \( G' \).

Next, observe that \( \alpha(G) = \alpha(G') + k \) and \( \alpha_u(G') + k = \alpha_u(G) \) for any vertex \( u \in G' \). Since for any \( y \geq x \) and \( k \geq 0 \), it holds that \( \frac{y}{x} \geq \frac{y+k}{x+k} \), it is straightforward to verify that
\begin{align*}
f\big(\alpha(G'), M',& \{\alpha_u(G') : u \in G'\}\big) \geq \\
&f\big(\alpha(G), M, \{\alpha_u(G) : u \in G\}\big),
\end{align*}
and the result follows.
\end{IEEEproof}

We conclude this section with a simple upper bound on $n^\bx_{G,q}(M,d)$ that uses the notion of graph coloring.
\begin{theorem} [Upper bound on $n^\bx_{G,q}(M,d))$]
For any graph $G$ with vertex set of size $M$, any distance $d$, and any $q$, $$n^\bx_{G,q}(M,d) \leq n^\bx_q(\chi(G),d) \leq n_q(\chi(G),d),$$ where
$\chi(G)$ is the chromatic number of $G$. 
\end{theorem}

\begin{IEEEproof}
Consider any vertex coloring of $G$ of size $\chi(G)$.
Let $\cC^\bx$ be an $(n,\chi(G),d)_q$ box code.
In particular, $\cC^\bx$ has $\chi(G)$ codewords, one distinct codeword for every color class in $G$.
To see that 
$n^\bx_{G,q}(M,d) \leq n^\bx_q(\chi(G),d)$, notice that $\cC^\bx$ can be extended to an $(n,M,d)_q$ code $\cC_G^\bx$ for $G$ by assigning identical codewords to all vertices in $G$ in the same color class; where each color class in $G$ is assigned the corresponding codeword from $\cC^\bx$.
In $\cC_G^\bx$, all pairs of codewords corresponding to an edge in $G$ are of distance at least $d$.
\end{IEEEproof}

\bibliographystyle{IEEEtranS}
\bibliography{allbib}

\begin{thebibliography}{10}
\providecommand{\url}[1]{#1}
\csname url@samestyle\endcsname
\providecommand{\newblock}{\relax}
\providecommand{\bibinfo}[2]{#2}
\providecommand{\BIBentrySTDinterwordspacing}{\spaceskip=0pt\relax}
\providecommand{\BIBentryALTinterwordstretchfactor}{4}
\providecommand{\BIBentryALTinterwordspacing}{\spaceskip=\fontdimen2\font plus
\BIBentryALTinterwordstretchfactor\fontdimen3\font minus
  \fontdimen4\font\relax}
\providecommand{\BIBforeignlanguage}[2]{{%
\expandafter\ifx\csname l@#1\endcsname\relax
\typeout{** WARNING: IEEEtranS.bst: No hyphenation pattern has been}%
\typeout{** loaded for the language `#1'. Using the pattern for}%
\typeout{** the default language instead.}%
\else
\language=\csname l@#1\endcsname
\fi
#2}}
\providecommand{\BIBdecl}{\relax}
\BIBdecl

\bibitem{AhlAydKha01}
R.~Ahlswede, H.~K. Aydinian, and L.~H. Khachatrian, ``On perfect codes and
  related concepts,'' \emph{Designs, Codes and Cryptography}, vol.~22, no.~3,
  pp. 221--237, Jan. 2001.

\bibitem{AhlKha98}
R.~Ahlswede and L.~H. Khachatrian, ``The diametric theorem in {H}amming spaces
  --- optimal anticodes,'' \emph{Advances in Applied Mathematics}, vol.~20, pp.
  429--449, 1998.

\bibitem{Alo23}
N.~Alon, ``On bipartite coverings of graphs and multigraphs,'' \emph{arXiv
  preprint arXiv:2307.16784}, 2023.

\bibitem{BorEtzRot24}
A.~Boruchovsky, T.~Etzion, and R.~M. Roth, ``On nearly perfect covering
  codes,'' \emph{arXiv preprint arXiv:2405.00258}, 2024.

\bibitem{EtzVar94}
T.~Etzion and A.~Vardy, ``Perfect binary codes: constructions, properties and
  enumeration,'' \emph{IEEE Trans.~Inform.~Theory}, vol.~40, no.~3, pp.
  754--763, May 1994.

\bibitem{Han64}
G.~Hansel, ``Nombre minimal de contacts de fermeture n{\'e}cessaires pour
  r{\'e}aliser une fonction bool{\'e}enne sym{\'e}trique de {$n$} variables,''
  \emph{C. R. Acad. Sci. Paris}, vol. 258, pp. 6037--6040, 1964.

\bibitem{KatSze67}
G.~Katona and E.~Szemer{\'e}di, ``On a problem of graph theory,'' \emph{Studia
  Sci. Math. Hungar.}, vol.~2, pp. 23--28, 1967.

\bibitem{KimLee23}
J.~Kim and H.~Lee, ``Covering multigraphs with bipartite graphs,'' \emph{arXiv
  preprint arXiv:2304.11691}, 2023.

\bibitem{MacSlo78}
F.~J. MacWilliams and N.~J.~A. Sloane, \emph{The Theory of Error-Correcting
  Codes}.\hskip 1em plus 0.5em minus 0.4em\relax North-Holland, 1978.

\bibitem{ShiWanAnKim24}
M.~Shi, X.~Wang, J.~An, and J.-L. Kim, ``Log-concave sequences in coding
  theory,'' \emph{arXiv preprint arXiv:2410.04412}, 2024.

\bibitem{turan1941external}
P.~Tur{\'a}n, ``On an external problem in graph theory,'' \emph{Mat. Fiz.
  Lapok}, vol.~48, pp. 436--452, 1941.

\bibitem{ulukus2015energy}
S.~Ulukus, A.~Yener, E.~Erkip, O.~Simeone, M.~Zorzi, P.~Grover, and K.~Huang,
  ``Energy harvesting wireless communications: A review of recent advances,''
  \emph{IEEE Journal on Selected Areas in Communications}, vol.~33, no.~3, pp.
  360--381, 2015.

\bibitem{Wee88}
G.~J.~M. {van Wee}, ``Improved sphere bounds on the covering radius of codes,''
  \emph{IEEE Trans.~Inform.~Theory}, vol.~34, no.~2, pp. 237--245, Mar. 1988.

\end{thebibliography}

\end{document}